\documentclass[prd,amsmath,aps,floats,amssymb, floatfix, superscriptaddress,preprintnumbers,twocolumn]{revtex4-1} 
\usepackage{aas_macros}
\usepackage[english]{babel}
\usepackage{amsmath}
\usepackage{graphicx}
\usepackage[colorlinks=true, allcolors=blue]{hyperref}

\begin{document}

\title{Results and forecasts on cosmic inflation from weak lensing}
\author{Agn\`es Fert\'e} 
\affiliation{SLAC National Accelerator Laboratory, Menlo Park, CA 94025, USA}
\author{Kevin Hong}
\affiliation{SLAC National Accelerator Laboratory, Menlo Park, CA 94025, USA}
\affiliation{Department of Physics \& Astronomy, 430 Portola Plaza, University of California, Los Angeles, CA 90095, USA}

\begin{abstract} 
We highlight the role of weak lensing measurements from current and upcoming stage-IV imaging surveys in the search for cosmic inflation, specifically in measuring the scalar spectral index $n_s$.  
To do so, we combine the Dark Energy Survey 3 years of observation weak lensing and clustering data with Bicep/\textit{Keck}, \textit{Planck} and Sloan Digital Sky Survey data in $r\Lambda$CDM where $r$ is the tensor-to-scalar ratio.
While there is no significant improvement in constraining power, we obtain a 1$\sigma$ shift on $n_s$. Additionally, we forecast a weak lensing and clustering data vector from the 10-year Legacy Survey of Space and Time by the Vera C. Rubin Observatory and show its combination with current data would improve their $n_s$ constraints by 25$\%$ in $r\Lambda$CDM.
\end{abstract}

\maketitle

\section{Introduction}

In the $\Lambda$CDM model, scalar perturbations of the metric have been evolving since cosmic inflation, sourcing the large-scale structures in the recent Universe.
Their primordial power spectrum $P_{\mathcal{S}}(k)$ follows: 
\begin{equation}
    P_{\mathcal{S}}(k) = A_s \left ( \frac{k}{k_s} \right )^{n_s -1},
\end{equation}
with $A_s$ the amplitude of scalar pertubation, $k$ the wavenumber, $k_s$ the wavenumber at a pivot scale, set to 0.05 $\rm{Mpc}^{-1}$ in this analysis, and $n_s$ the scalar spectral index.

The implications of Cosmic Microwave Background (CMB) measurements for inflationary search were introduced in \cite{knox1,knox2,cmb_inflation}.
And indeed, since then, one of the successes of the last generation of CMB experiments was measuring $n_s$ different from unity at high significance (8$\sigma$ in \cite{planck18_main}) thus excluding a scale-invariant primordial power spectrum, a key step towards establishing inflation.
Now and over the coming decade, CMB polarization experiments aim at detecting large scale $B$-modes to constrain the tensor-to-scalar ratio $r$, the energy scale of inflation.
Today, the Bicek/\textit{Keck} experiment in combination with CMB \textit{Planck} and baryonic acoustic oscillations (BAO) measurements from 6dFGS, MGS and BOSS DR12 data constrain $r$ to be below 0.036 at 95\% confidence \cite{bk_18}.
In the future, the \textit{Litebird} satellite \cite{litebird}, the Simons Observatory \cite{simonsobservatory} and CMB-S4 \cite{cmbs4} ground-based experiments will aim at attaining $\sigma(r) \sim 10^{-3}$. 

In parallel, a new generation of photometric galaxy surveys will soon start mapping galaxies to further test the $\Lambda$CDM model with the goal of understanding the origin of the current cosmic acceleration. 
On the ground, the Vera C. Rubin Observatory will produce the Legacy Survey of Space and Time (LSST) \cite{lsst_whitepaper,lsst_srd}, a 10-year imaging survey of half the celestial sphere, while the Euclid \cite{euclid_whitepaper} and Roman \cite{roman_whitepaper} satellites will image galaxies from space. 
One of their main objectives is to probe dark energy and modified gravity, through the evolution of the background and structures in the recent Universe through various observables \cite{desc_srd,lsst_mg,euclid_forecast,roman_forecast}.
Weak gravitational lensing is especially promising as being one of the few unbiased probes of dark matter distribution and having been successfully used for precision cosmology with stage-III experiments \cite{desy3,hsc32pt,kids1000}.
Weak lensing is mostly sensitive to the energy density of matter $\Omega_m$ and the variance of matter fluctuations $\sigma_8$, however with improved measurements from stage-IV surveys (as defined in \cite{stage4}), weak lensing will become more sensitive to other properties of the matter power spectrum: in this paper, we investigate the role of weak lensing in inflationary search, especially through its sensitivity to the scalar spectral index $n_s$.
Weak lensing indeed brings complementary information from CMB, by accessing different modes, in the range $k \sim [0.1,5] h$/Mpc, compared to the range accessed by the CMB $k \sim [10^{-4},10^{-1}] h$/Mpc.
We note that in parallel, galaxy surveys will also aim at detecting primordial non-Gaussianities, through galaxy clustering as forecasted in \cite{fnl_forecast,fnl_forecast_euclidska,spherex}, as well as through weak lensing \cite{fnl_shear} and alignment of galaxies \cite{fnl_ia}. 

There are now a few indications that weak lensing could bring promising improvements on $n_s$ constraints. In \cite{Gatti_2020}, the second moment of the mass map from the Dark Energy Survey shows sensitivity to $n_s$. Additionally, \cite{Prat_2023,Tan_2023} indicate significant constraints on $n_s$ with stage-IV surveys, although they are probably partly informed by prior choices.
Weak lensing was also used in \cite{Chandra_2022} to determine the sensitivity of \textit{Euclid}-like surveys in detecting specific features arising from inflation. 
Furthermore, \cite{Huang_2012} shows that future spectroscopic surveys such as \textit{Euclid} will increase the constraints in the $n_s$ direction by close to a factor of 2 in $r\Lambda$CDM.
However, this analysis did not consider weak lensing, so we complete the picture in the present paper by considering information from photometric surveys, specifically the Dark Energy Survey (DES) and the future LSST.
To do so, we first infer cosmology in $\Lambda$CDM  and $r\Lambda$CDM using data from the DES 3 years of observation (DES Y3) and second, from our predicted 10-year LSST data vector. We describe both datasets in Sec.~\ref{sec:forecast} and detail the other likelihoods used as well as our parameter estimation approach in Sec.~\ref{sec:analysis}. We show our results in $\Lambda$CDM and $r\Lambda$CDM models in Sec.~\ref{sec:implication}.
We finally conclude in Sec.~\ref{sec:conclusion} with outlooks on weak lensing's role in inflation search. 

\section{Analysis}
\subsection{DES Y3 and predicted LSST Y10 weak lensing and clustering}
\label{sec:forecast}

To quantify weak lensing contributions to constraints on inflation, we choose to combine information from weak lensing and clustering in order to pin down systematics such as intrinsic alignment and galaxy bias as done in \cite{desy3,kids1000,hsc32pt}.
Their statistics are summarized in the form of three correlation functions in tomographic bins (referred to as 3x2pt): cosmic shear $\xi_\pm(\theta)$ corresponding to the correlations of galaxy shapes, galaxy-galaxy lensing $\gamma_t(\theta)$, the tangential shear of background galaxies around lens galaxies, and finally clustering $w(\theta)$ corresponding to the correlation of lens galaxy positions.
We use DES Y3 3x2pt along with the modeling choices and angular scale cuts used in DES Y3 cosmological analysis in \cite{desy3}.
We tested that adding the DES Y3 shear ratio likelihood from \cite{desy3_sr} did not change the results. 

To forecast weak lensing from stage-IV surveys, we simulate a data vector from the 10 years of LSST (hereafter LSST Y10). 
For simplicity, in this case, we choose to use angular power spectra in harmonic space $C_{\ell}^{ab}$ (with $a$ and $b$ either the convergence field $\kappa$ or the density $\delta$) as our summary statistics. 
We closely follow the choices made in \cite{desc_srd} which we will refer to as the SRD (the LSST-Dark Energy Survey Science Collaboration Science Requirement Document) with small changes for more realistic forecasts which we describe below.

Regarding LSST Y10 observations, we set the observed sky fraction used to create the source and lens samples to be 35$\%$ of the celestial sphere. 
The  redshift distribution $n(z)$ of both samples is described by a Smail distribution \textit{i.e.}:
\begin{equation}
    n(z) = z^{\alpha} e^{(-z/z_0)^{\beta}}.
    \label{eq:smail}
\end{equation}
Parameters $\alpha$, $\beta$ and $z_0$ of this distribution along with the number of redshift bins, effective number density and shape noise are summarized in Table \ref{tab:datachoice}, following the SRD.

\begin{table}
\begin{tabular}{|c|c|c|}
     \hline 
     Parameters & Source sample & Lens sample  \\
     \hline
     \hline
     $\alpha$      & 2             & 2 \\  
     $z_0$         & 0.11            & 0.28 \\  
     $\beta$       & 0.68            & 0.9 \\  
     Number of redshift bins & 5 & 10 \\
     $n_{\textrm{eff}}$ (in $\textrm{arcmin}^{-2}$) & 27 & 48 \\
     $\sigma_{\epsilon}$ & 0.26 & -- \\
     \hline
     \end{tabular}

\caption{Parameters used to model the redshift distribution of the lens and galaxy samples (see Eq. \ref{eq:smail}), along with the number of redshift bins, effective number density $n_{\textrm{eff}}$ and shape noise $\sigma_{\epsilon}$ used to simulate a LSST Year 10 weak lensing and clustering data vector, following \cite{desc_srd}.}
\label{tab:datachoice}
\end{table}

We list below the choices made to model the LSST Y10 $C_{\ell}$ data vector: 
\begin{itemize}
    \item The matter power spectrum is computed using \textsc{camb} \cite{Lewis:2002ah,Lewis:1999bs,Howlett:2012mh,camb_notes}. As the SRD uses scales down to $\ell_{\textrm{shear}} = 3000$, we decided to add a non-linear prescription with baryonic feedback from \textsc{HMCode-2020} \cite{Mead_2020}, to model the small angular scales more realistically. We set $\log_{10}(T_{\mathrm{AGN}}/K) = 7.8$ inside the range recommended in \cite{Mead_2020}, with $T_{\mathrm{AGN}}$ corresponding to the strength of Active Galactic Nuclei feedback in simulations.  
    \item The intrinsic alignment (IA) of galaxies is modeled using the non-linear alignment model \cite{Bridle_2007} such that the IA contributions to cosmic shear are linearly related to the non-linear matter power spectrum, where the amplitude of IA has (1+$z$) redshift dependence as used in DES Year 1 \cite{desy1_methods}. 
    \item Similarly to the SRD, we adopt a linear galaxy bias model parametrized by a bias parameter per redshift bin.  
\end{itemize}

\begin{table}
\begin{tabular}{|c|c|c|c|c|c|c|c|c|c|c|}
\hline
    Lens Bin & 1 & 2 & 3 & 4 & 5 & 6 & 7 & 8 & 9 & 10  \\
    $\langle z_{\ell} \rangle$ & 0.27 & 0.46 &  0.61 & 0.74 & 0.88 & 1.03 & 1.20 & 1.42 & 1.73 & 2.45 \\
    $\ell_{\textrm{max}}$ & 227 & 370 & 466 & 550 & 629 &  708 & 791 & 884 & 1000 & 1210 \\
\hline
\end{tabular}
\caption{Harmonic-space scale cuts adopted in the present analysis following \cite{desc_srd}, \textit{i.e.} $\ell_{\textrm{max}} = 3000$ for shear and $k_{\textrm{max}} = 0.3 h/\textrm{Mpc}$ for correlations involving the lens sample. The total number of data points used is 513. }
\label{tab:scalecuts}
\end{table}

We use \textsc{CosmoSIS} \cite{Zuntz_2015} to model and analyze DES Y3 and our LSST Y10-like data vector. 
We thus theoretically predict the expected $C_{\ell}$, with a Gaussian covariance matrix computed within \textsc{CosmoSIS}.
Weak lensing analyses such as \cite{desy3} remove measurements, typically at small angular scales, where the modeling is uncertain. 
We do similarly and follow the guidelines from the SRD, thus using $\ell_{\textrm{max}} = 3000$ for weak lensing, and $k_{\textrm{max}} = 0.3 h/\textrm{Mpc}$ for the clustering part of the data vector. We translate this value into corresponding $\ell_{\textrm{max}}$ for each lens redshift bins as shown in Table~\ref{tab:scalecuts}.

\begin{figure}
    \centering
    \hbox{\hspace{-0.3cm}\includegraphics[width=0.5\textwidth,keepaspectratio]{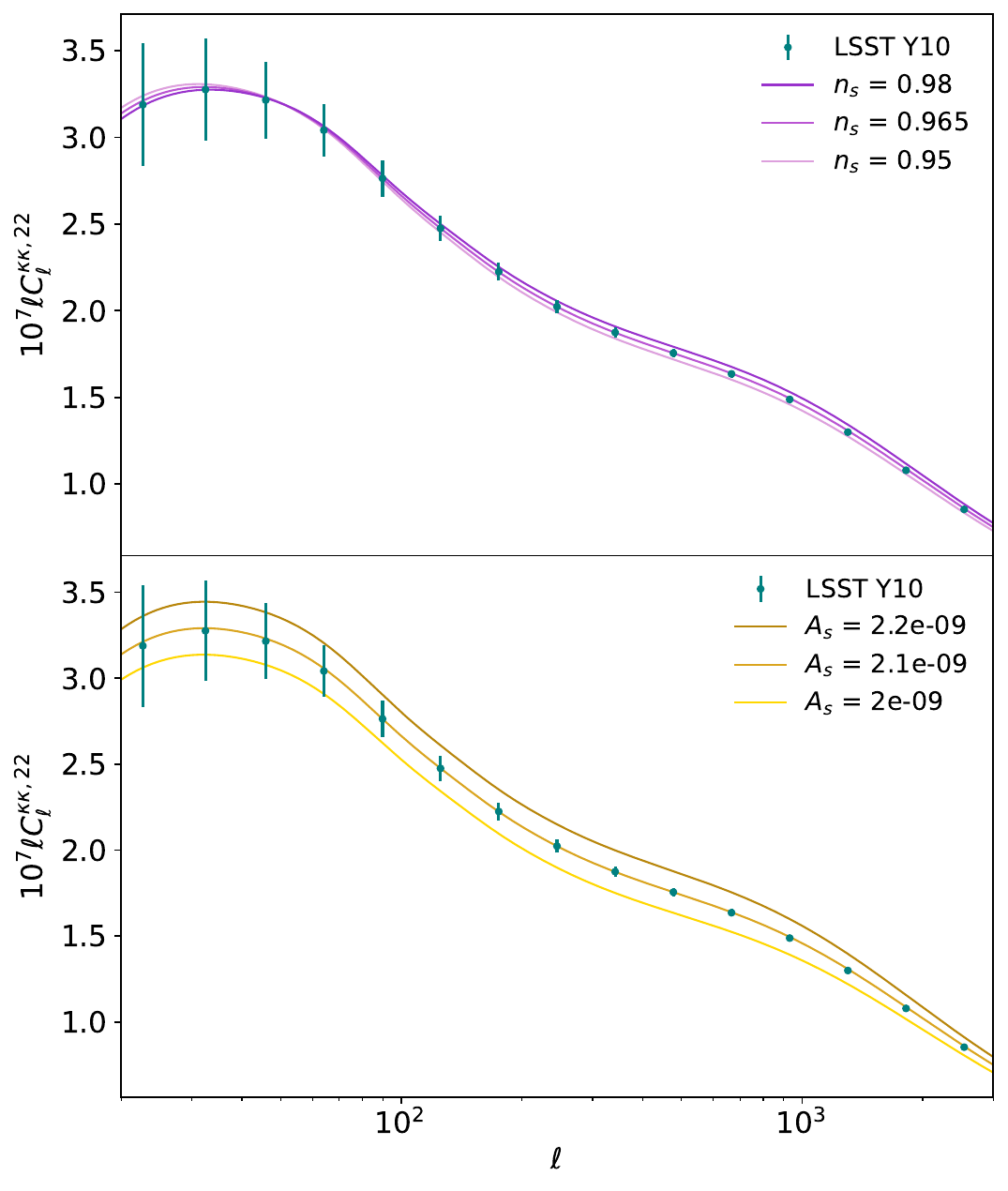}}
    \caption{Theoretical predictions of the shear angular power spectra in redshift bin 2 of the LSST Y10 source $n(z)$ for three values of the scalar spectral index $n_s$ in the top panel, and three values of the amplitude of scalar perturbations $A_s$ in the lower panel, along with our forecasted LSST Year 10 shear data vector and error bars. }
    \label{fig:cellns}
\end{figure}

The effect of the scalar spectral index $n_s$ on shear power spectrum in redsfhit bin 2 is shown on the top panel of Fig. \ref{fig:cellns} along with the predicted data points and error bars from LSST Y10.
The tilt of the primordial power spectrum translates into a dampening at low-$\ell$ (of at most 3$\%$ for $n_s$ = 0.98) and a boost at higher $\ell$.
Given the shown error bars, we thus expect LSST Y10 weak lensing to have sensitivity to this parameter.
We however also show the effect of the amplitude of scalar perturbations $A_s$ on the lower panel of Fig.~\ref{fig:cellns}, which behaves similarly to $n_s$ on small angular scales, where the sensitivity is best.
This translates into a degeneracy between the two parameters which is shown in the contours from analyzing our LSST Y10 3x2pt data dector in $\Lambda$CDM in  Fig.~\ref{fig:lssty10cosmo}.
For similar reasons, the baryonic feedback parameter $T_{\mathrm{AGN}}$ is degenerate with $n_s$ while the energy density of baryons $\Omega_b$ and the Hubble parameter $H_0$ are anti-correlated with $n_s$.
We therefore need to combine LSST Y10 with other data such as \textit{Planck} temperature and polarization power spectra to break such degeneracy, which we do in the following section.

\subsection{Other datasets and likelihood analysis}
\label{sec:analysis}
We adopt a Bayesian approach, where we sample the posterior using the \textsc{nautilus} importance nested sampler \cite{lange2023nautilus} within \textsc{CosmoSIS}.
The parameter estimation is made in $\Lambda$CDM to forecast the sensitivity of LSST Y10 on $n_s$ and in $r\Lambda$CDM to forecast implications of DES Y3 and LSST Y10 for cosmic inflation.

Although our LSST Y10 data vector is a theoretical prediction, we combine it with current real CMB and BAO data while of course measurements of such observables will also become more precise in the coming decade.
We indeed want to show results in the current experimental landscape first, using weak lensing data from DES Y3 in $r\Lambda$CDM and then show how solely improving DES Y3 to LSST Y10 would translate into inflation constraints.
We thus use \textit{Planck} 2018 temperature, polarization $E$-modes and lensing potential $\phi$ power spectra in the form of the $TT$, $TE$, $EE$ \textit{lite} high-$\ell$, $EE$ and $TT$ low-$\ell$ as well as lensing likelihoods, to add information on the cosmological parameters \cite{planck18_main}. 
We refer to this combination as TTTEEE+low-$\ell$+lensing. 
Additionally, to inform the geometry of the Universe and break the degeneracy between $\sigma_8$, $\Omega_m$ and the Hubble parameter $H_0$, we add Baryonic Acoustic Oscillations measurements from Sloan Digital Sky Survey. They specifically include likelihoods on distance measurements from the Main Galaxy Sample (MGS) \cite{mgs_bao}, the Baryon Oscillation Spectroscopic Survey (BOSS) DR12 \cite{boss_dr12} re-analyzed in \cite{sdss} and extended Baryon Oscillation Spectroscopic Survey (eBOSS) DR16 measurements from luminous red galaxies \cite{eboss_lrg1,eboss_lrg2}, emission line galaxies \cite{elg}, quasars \cite{qso1,qso2} and Ly-$\alpha$ forest \cite{lyalpha}.

\begin{table}
\begin{tabular}{|c|c|c}
\hline
    Parameters & Priors \\
\hline
\hline
     \multicolumn{2}{|c|}{Cosmology} \\
\hline    
    $A_s$ & [0.5,5] $\times 10^{-9}$ \\
    $n_s$ & [0.88,1] \\
    $\Omega_m$ & [0.1,0.7] \\
    $\Omega_b$ & [0.03,0.07] \\
    $h_0$ & [0.55,0.9] \\
    $r$ (in $r\Lambda$CDM) & [0,0.2] \\
\hline
\hline
 \multicolumn{2}{|c|}{DES Y3} \\
 \hline 
 \multicolumn{2}{|c|}{See Table 1 in \cite{desy3}}\\
\hline
\hline
     \multicolumn{2}{|c|}{Forecast LSST Y10} \\
\hline    
    $\log(T_{\rm{AGN}})$ & [7.7,8.0] \\
    $A_{\rm{IA}}$ & [-5,5] \\
    $\alpha$ & [-5,5] \\
    $m_i, i \in [1,5]$ & [-0.005,0.005] \\
    $\Delta z_{s}^{i}, i \in [1,5] $ & [-0.01,0.01] \\
    $\Delta z_{\ell}^{i}, i \in [1,10]$ & [-0.01,0.01] \\
    $b_i, i \in [1,10]$ & [1.9,2.1] \\
\hline
\hline
     \multicolumn{2}{|c|}{\textit{Planck}} \\
\hline    
    $\tau$ & [0.01,0.8] \\
    $A_{\rm{Planck}}$ & $\mathcal{G}$(1,0.0025)\\
\hline
\end{tabular}
\caption{ Priors on parameters used in the parameter estimation, where brackets indicate flat priors while $\mathcal{G}(m,\sigma)$ indicates a Gaussian prior of mean $m$ and standard deviation $\sigma$. }
\label{tab:priors}
\end{table}

In $r\Lambda$CDM, we additionally include the likelihood on the tensor-to-scalar ratio $r$ from Bicep/\textit{Keck} $B$-modes power spectrum measurements from \cite{bk_18} (hereafter BK18). 
We note that Fig.~\ref{fig:rns} shows in coral the combination of BK18, Planck TTTEEE+low-$\ell$+lensing and SDSS including eBOSS DR16, while BK18 analysis in \cite{bk_18} uses 6dFGS, MGS and BOSS DR12. 
The shift in $n_s$ caused by this update in the BAO measurement is not significant (0.07$\sigma$).

Table~\ref{tab:priors} summarizes the parameters varied and their corresponding priors, including both cosmological and nuisance parameters. 
We use GetDist \cite{getdist} to quote constraints on parameters as the mean and 68$\%$ credible intervals in one dimension, and to show the 68$\%$ and 95$\%$ credible regions in two dimensions.

\section{Results, Forecasts and Implications for inflation}
\label{sec:implication}

\begin{figure}
    \centering
    {\hspace{-0.4cm}\includegraphics[width=0.5\textwidth]{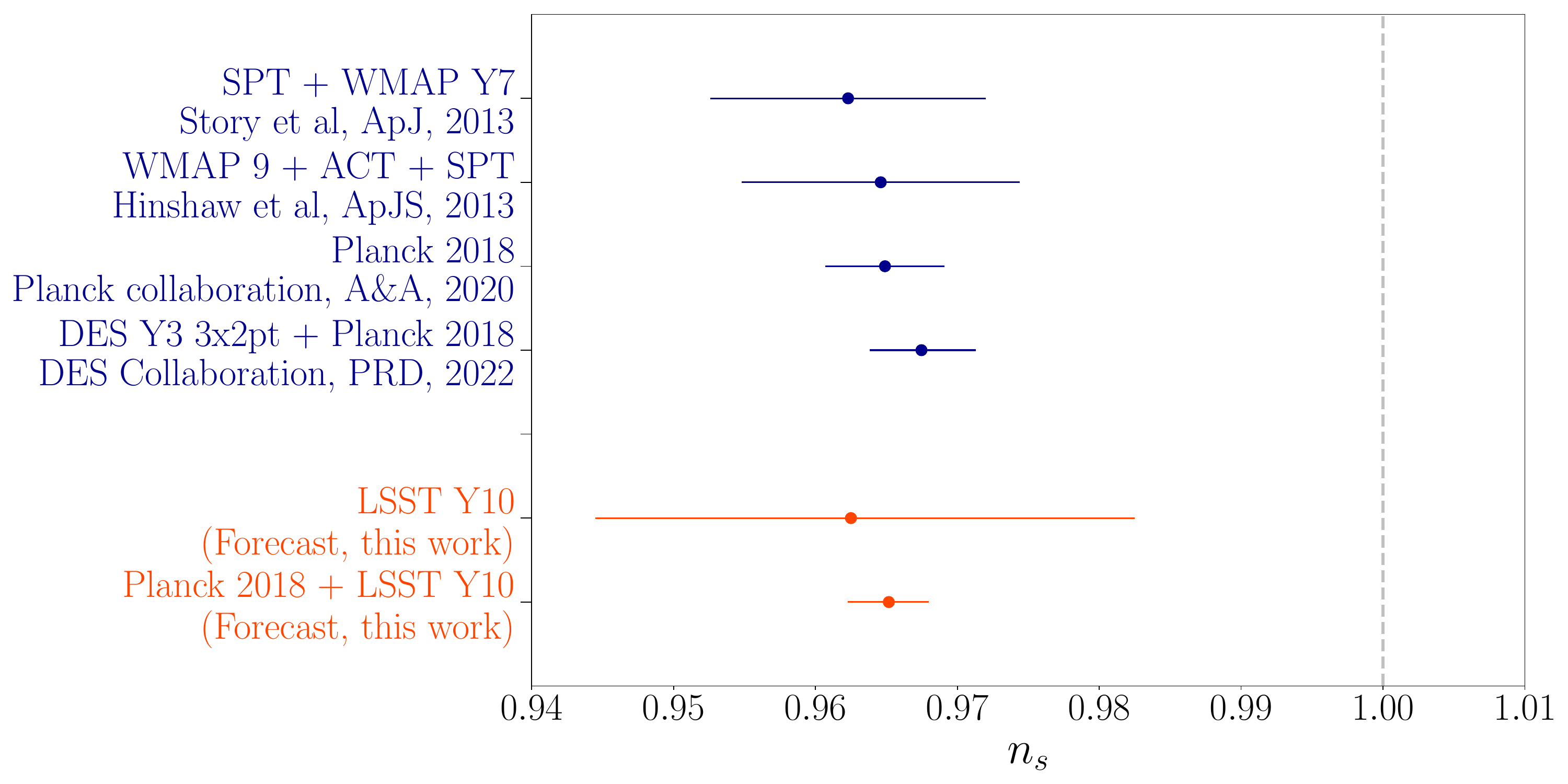}}
    \caption{Past and current measurements of the scalar spectral index $n_s$ in blue along with the predictions from our forecasted LSST Year 10 weak lensing and clustering data vector, alone and in combination with \textit{Planck} TTTEEE, low-$\ell$ and lensing likelihoods in orange. }
    \label{fig:ns1d}
\end{figure}

In Fig.~\ref{fig:ns1d}, we summarize the present status of $n_s$ measurements in $\Lambda$CDM, quoting results from WMAP data, the first rejection of $n_s = 1$ at high significance in \cite{story_measurement_2013,Hinshaw_2013}, and the tightest current measurements from the \textit{Planck} satellite \cite{planck18_main} as well as the results from DES Y3 combined with \textit{Planck} (which does not include CMB lensing) as published in \cite{desy3}. 
We then report the mean and 68$\%$ credible interval on $n_s$ we obtain from our analysis of LSST Y10 3x2pt alone and LSST Y10 3x2pt combined with \textit{Planck} TTTEEE+low-$\ell$+lensing in orange. 
While LSST Y10 by itself does not result in competitive results on the spectral index, the forecast shows a 30 $\%$ improvement on $n_s$ from adding LSST Y10 to \textit{Planck} compared to \textit{Planck} alone.
Although the gain in constraining power on $n_s$ expected from spectroscopic surveys is greater \cite{Huang_2012}, weak lensing and clustering prove to be useful additions as probes of the matter power spectrum.
In Fig.~\ref{fig:lssty10cosmo} in the appendix, we show the predicted constraints on cosmology from LSST Y10 3x2pt in red as well as the combination with \textit{Planck} in dark blue.

\begin{figure}
    \centering
 \hbox{\hspace{-0.3cm}\includegraphics[width=0.55\textwidth]{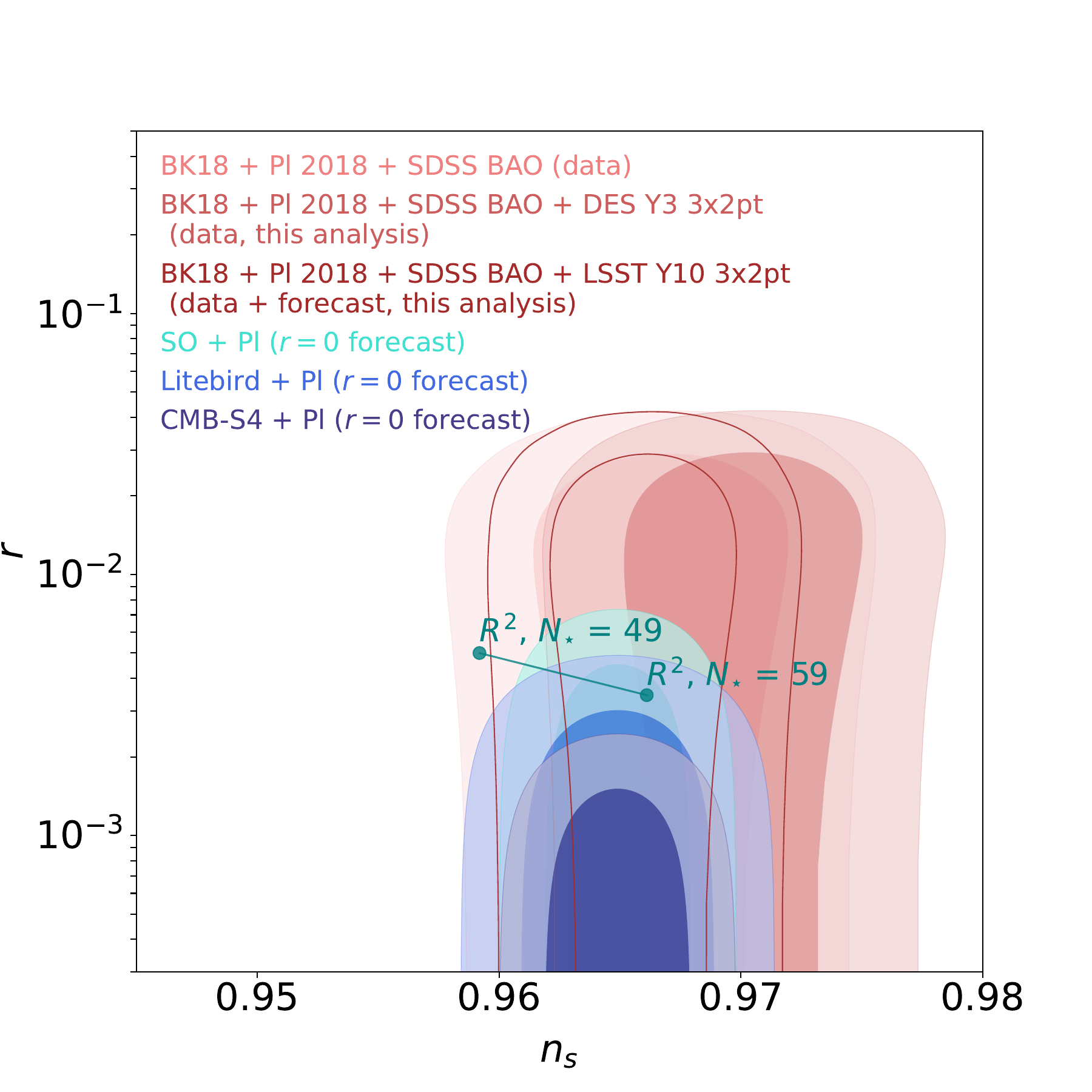}}
    \caption{Current constraints and forecasts on the tensor-to-scalar ratio $r$ and the scalar spectral index $n_s$ using the likelihood on $r$ from Bicep/\textit{Keck} in addition to \textit{Planck} 2018 TTTEEE, low-$\ell$ and lensing likelihoods, BAO measurements from SDSS, in salmon. We add DES Y3 weak lensing and clustering data to this fiducial data combination, and show the results in red. Forecasts from replacing DES Y3 by our LSST Y10-like weak lensing and clustering  data vector is shown in brown. We also show for reference forecasts for future Stage-IV CMB experiments: Simons Observatory (in turquoise), \textit{Litebird} (in blue) and CMB-S4 (in dark blue), with the prediction from Starobinsky inflationary model (referred to as $R^2$ model) for e-fold $N_{\star}$ between 49 and 59 overlaid in teal.  }
    \label{fig:rns}
\end{figure}

We now turn to the implications of such improvements on inflationary models, in $r\Lambda$CDM. 
The results from our parameter inference are summarized in Fig.~\ref{fig:rns} where we show constraints in the ($r$,$n_s$) plane, using the BK18 likelihood on the tensor-to-scalar ratio in addition to \textit{Planck} TTTEEE+low-$\ell$+lensing and SDSS BAO measurements in salmon.
As a reference, we also show predictions from the Starobinsky model \cite{r2model1,r2model2} for e-folds $N_{\star}$ between 49 and 59 in teal.
As nicely summarized in \cite{Hinshaw_2013}, $r$ and $n_s$ are indeed simply related to the number of e-fold in the super-horizon limit, following \cite{r2model3}. In particular: $n_s - 1 = - 2/N_{\star}$.

First, adding DES Y3 3x2pt results in the red contour on ($r$,$n_s$). 
There is virtually no improvement on $n_s$ but strikingly, the contours are shifted by 1$\sigma$ to higher values of $n_s$, indicating that any e-folds $N_{\star}$ below 55 are rejected at more than 2$\sigma$.
We believe this shift to be caused by the slight tension between \textit{Planck} and DES Y3 3x2pt, where DES Y3 pulls the results towards lower values of $\Omega_m$ which in turn translates into higher values of $n_s$ (given the slight anti-correlation in \textit{Planck}'s ($\Omega_m$,$n_s$) plane).

We then switch DES Y3 to our predicted LSST Y10 3x2pt data vector and report the result in brown, indicating that the 68$\%$ credible interval on $n_s$ would in this case be improved by 25$\%$.
We note that our predicted LSST Y10 3x2pt data vector was computed for a value of $n_s$ equal to its mean measured by \textit{Planck} 2018.
The addition of LSST Y10 3x2pt will thus help test the $R^2$ model more strongly.

In the future, stage-IV CMB experiments dedicated to inflation search will aim at $\sigma(r) \sim 10^{-3}$.
The combination of their polarization power spectra with \textit{Planck} will also tighten the constrains on $n_s$.
As a reference we show forecasts for Simons Observatory (SO) in cyan, \textit{Litebird} in light blue and CMB-S4 in blue, taken from their forecast papers \cite{simonsobservatory,litebird,cmbs4}.
In future work, we will assess the expected improvements on $n_s$, and therefore on $R^2$ model constraints, from combining these experiments with future weak lensing (as shown here) and spectroscopic clustering (as shown in \cite{Huang_2012}) measurements.

\begin{table}
\begin{tabular}{|c|c|}
     \hline 
     $r\Lambda$CDM & $n_s$   \\
     \hline
     \hline
      Fiducial &  $0.9668 ^{+0.0037}_{-0.0035}  $ \\
      Fiducial + DES Y3 3x2pt  & $0.9702 ^{+0.0034}_{-0.0035}$ \\
      Fiducial + forecast LSST Y10 3x2pt & $0.9660^{+ 0.0028}_{- 0.0027}$ \\      
     \hline
     \end{tabular}

\caption{Mean and 68$\%$ credible interval on the scalar spectral index $n_s$ in $r\Lambda$CDM from the fiducial combination of current data (\textit{i.e.} BK18 + \textit{Planck} 2018 (TTTEEE+low-$\ell$+lensing) + SDSS BAO (MGS,eBOSS DR16)) and combination with weak lensing and clustering data.}
\label{tab:nsresults}
\end{table}

\section{Conclusion}
\label{sec:conclusion}

The detection of Cosmic Microwave Background $B$-modes on large scales is a great goal of modern cosmology as an awaited signal from cosmic inflation.
Experiments such as BICEP/\textit{Keck} have therefore been developed to enable the current tightest constraints on the tensor-to-scalar ratio $r$, with implications for inflation shown as constraints on $r$ and the scalar spectral index $n_s$ of scalar perturbations as shown in Fig.~\ref{fig:rns}.
In the coming decade, a new generation of CMB polarization experiments including Simons Observatory, \textit{Litebird}, CMB-S4 will aim at improving constraints in the $r$ direction, by attaining $\sigma(r) \sim 10^{-3}$.
However, we also need improvements in the $n_s$ direction to help further test inflationary models (such as $R^2$ model), in particular \cite{Huang_2012} already showed the power of spectroscopic measurements for surveys like \textit{Euclid} in improving $n_s$ constraints by a factor of 2.
In the present analysis, we complete the picture by showing the expected improvements from current and stage-IV weak lensing surveys in inflationary search and summarize our results in Table \ref{tab:nsresults}. 

The next steps in this direction include assessing weak lensing and clustering sensitivity to the running of the scalar index, $\alpha_s$, the derivative of $n_s$ to the wavenumber $k$ as well as including information from the mass and galaxy maps beyond two-point statistics.
Additionally, we show results with primordial and lensing CMB information from \textit{Planck} but further work will be needed to understand how future galaxy surveys will help stage-IV CMB experiments in improving $n_s$ constraints.
On a similar note, we show  in Fig.~\ref{fig:lsstnsprior} how $n_s$ priors informed by the CMB will impact cosmology from LSST Y10, the parameter $\sigma_8$ appearing unchanged.
To conclude, given the experimental landscape of the coming decade, we will want to combine results from both spectroscopic and weak lensing surveys with CMB polarization data to perform more complete inflation searches.

\acknowledgments
AF would like to thank Daniel Green for inputs on CMB-S4 forecasts and Judit Prat for discussions about LSST Y1 forecasts.

This work was supported in part by the U.S. Department of Energy, Office of Science, Office of Workforce Development for Teachers and Scientists (WDTS) under the Science Undergraduate Laboratory Internships Program (SULI).
Some of the computing for this project was performed on the Sherlock cluster.
We would like to thank Stanford University and the Stanford Research Computing Center for providing computational resources and support that contributed to these research results.
We thank the developers of \textsc{CosmoSIS} and modules therein, the parameter inference tool used in this analysis, available at this \href{https://cosmosis.readthedocs.io/en/latest/}{link}.

\appendix*
\section{LSST Y10 weak lensing and clustering forecast}
\label{app:cosmo}

\begin{figure*}
    \centering
\includegraphics[width=7.1in]{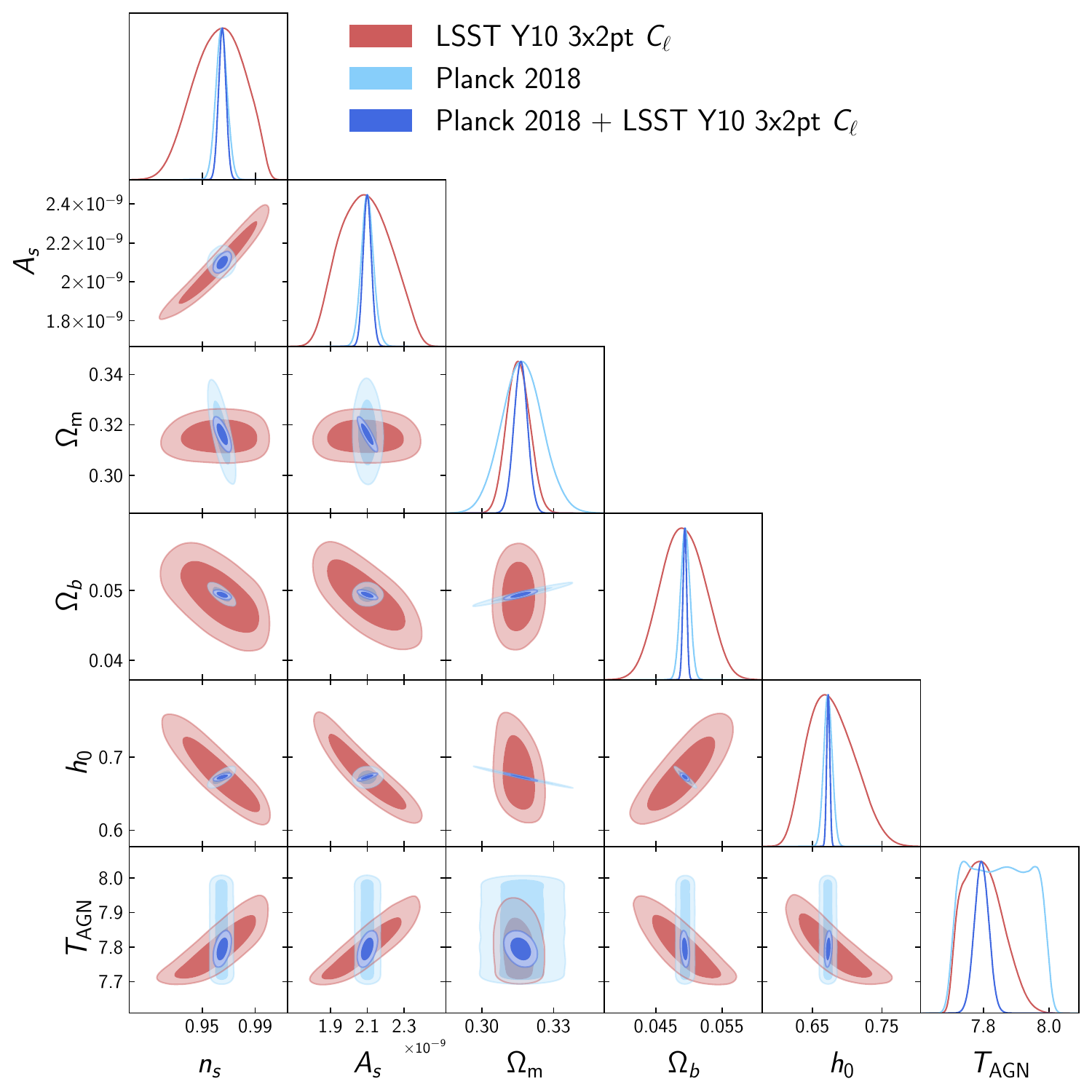}
    \caption{Credible regions on cosmological parameters from our predicted LSST Y10 weak lensing and clustering data vector in harmonic space in $\Lambda$CDM in red along with constraints on corresponding parameters from \textit{Planck} 2018 TTTEEE+low-$\ell$+lensing likelihoods in light blue and their combination with LSST Y10 in blue.  }
    \label{fig:lssty10cosmo}
\end{figure*}

We show in Fig.~\ref{fig:lssty10cosmo} the credible regions of cosmological parameters obtained by analyzing our predicted LSST Y10 3x2pt $C_{\ell}$ data vector in $\Lambda$CDM in red, from \textit{Planck} 2018 TTTEEE+low-$\ell$+lensing likelihoods in light blue and their combination with LSST Y10 in dark blue.
We thus show how LSST and \textit{Planck} combined together will improve cosmological constraints by breaking several degeneracies.  

\begin{figure*}
    \centering
\includegraphics[width=7.1in]{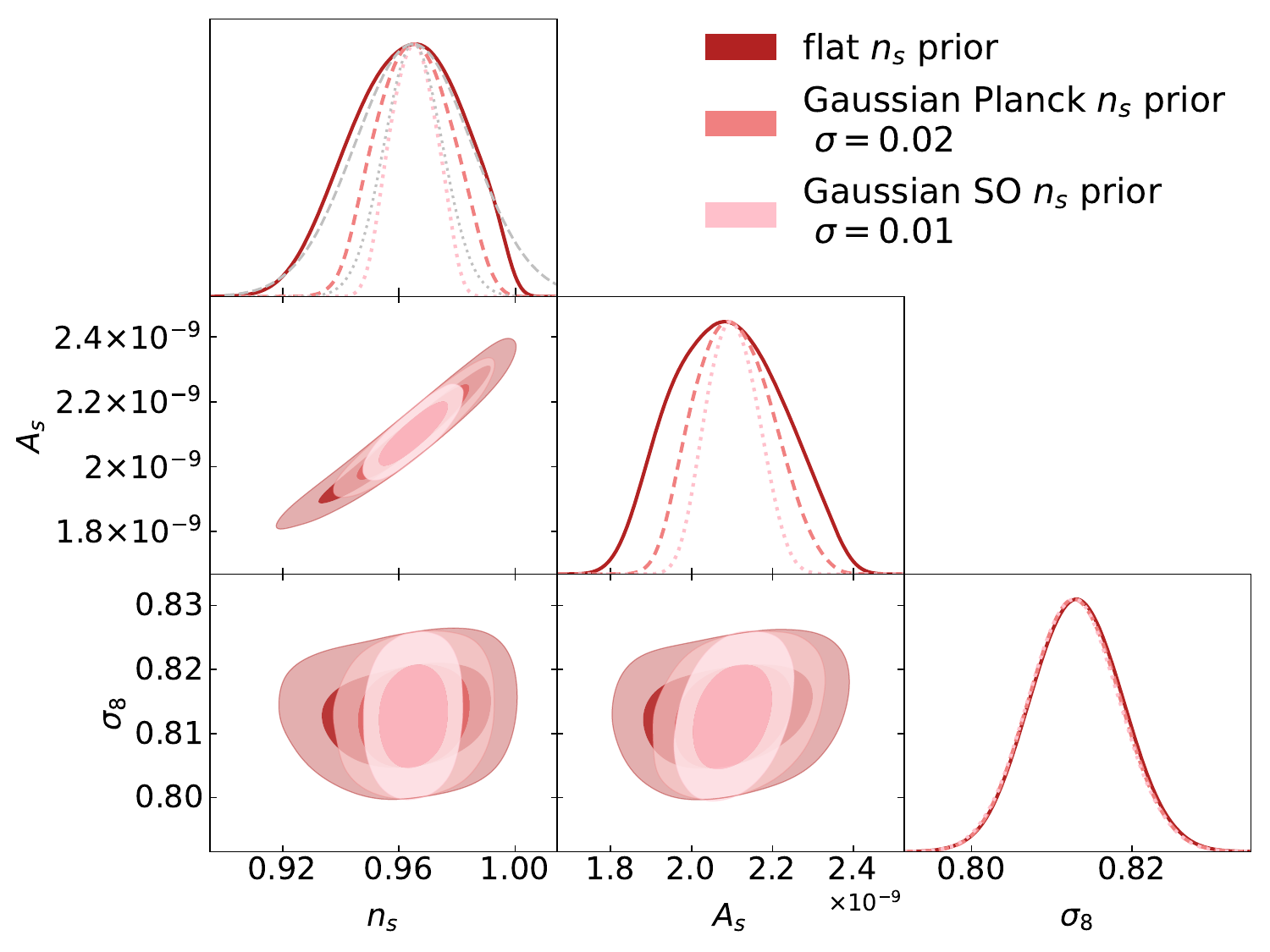}
    \caption{Credible regions on the scalar spectral index $n_s$ and amplitude $A_s$ of scalar pertubations and $\sigma_8$ from our predicted LSST Y10 weak lensing and clustering data vector in harmonic space in red. In coral, we show the results from using a Gaussian prior with a width 5 times larger than \textit{Planck} 2018 (\textit{i.e.} $\sigma = 0.02$) on $n_s$, the prior is shown as a dashed grey line on the 1D $n_s$ panel. We similarly show results from adding a Gaussian prior with a width 5 times larger than predicted from Simons Observatory (\textit{i.e.} $\sigma = 0.01$) in pink, the prior is shown as a dotted grey line on the 1D $n_s$ panel. }
    \label{fig:lsstnsprior}
\end{figure*}

In Fig.~\ref{fig:lsstnsprior}, we show forecasts on $A_s$,$n_s$ and $\sigma_8$ obtained analyzing LSST Y10 3x2pt using a flat wide prior on $n_s$ (shown in red), a Gaussian prior informed by \textit{Planck} 2018 (in coral) and SO (in pink). In both cases, the standard deviation of the Gaussian prior is 5 times the 68$\%$ credible interval from \textit{Planck}, and 5 times the expected uncertainty from SO (expected to be twice as small as \textit{Planck}). 

\bibliographystyle{ieeetr}
\bibliography{main_arxiv}

\begin{thebibliography}{10}

\bibitem{knox1}
L.~Knox and M.~S. Turner, ``Detectability of tensor perturbations through
  anisotropy of the cosmic background radiation,'' {\em Phys. Rev. Lett.},
  vol.~73, pp.~3347--3350, Dec 1994.

\bibitem{knox2}
L.~Knox, ``Determination of inflationary observables by cosmic microwave
  background anisotropy experiments,'' {\em Phys. Rev. D}, vol.~52,
  pp.~4307--4318, Oct 1995.

\bibitem{cmb_inflation}
S.~Dodelson, W.~H. Kinney, and E.~W. Kolb, ``Cosmic microwave background
  measurements can discriminate among inflation models,'' {\em Phys. Rev. D},
  vol.~56, pp.~3207--3215, Sep 1997.

\bibitem{planck18_main}
{Planck Collaboration}, {Aghanim, N.}, {Akrami, Y.}, {Ashdown, M.}, {Aumont,
  J.}, {Baccigalupi, C.}, {Ballardini, M.}, {Banday, A. J.}, {Barreiro, R. B.},
  {Bartolo, N.}, {Basak, S.}, {\em et~al.}, ``Planck 2018 results - vi.
  cosmological parameters,'' {\em A\&A}, vol.~641, p.~A6, 2020.

\bibitem{bk_18}
P.~A.~R. Ade, Z.~Ahmed, M.~Amiri, D.~Barkats, R.~B. Thakur, C.~A. Bischoff,
  D.~Beck, J.~J. Bock, H.~Boenish, E.~Bullock, {\em et~al.}, ``Improved
  constraints on primordial gravitational waves using planck, wmap, and
  bicep/keck observations through the 2018 observing season,'' {\em Phys. Rev.
  Lett.}, vol.~127, p.~151301, Oct 2021.

\bibitem{litebird}
{LiteBIRD Collaboration}, E.~Allys, K.~Arnold, J.~Aumont, R.~Aurlien,
  S.~Azzoni, C.~Baccigalupi, A.~J. Banday, R.~Banerji, R.~B. Barreiro,
  N.~Bartolo, {\em et~al.}, ``{Probing cosmic inflation with the LiteBIRD
  cosmic microwave background polarization survey},'' {\em Progress of
  Theoretical and Experimental Physics}, vol.~2023, p.~042F01, 11 2022.

\bibitem{simonsobservatory}
P.~Ade, J.~Aguirre, Z.~Ahmed, S.~Aiola, A.~Ali, D.~Alonso, M.~A. Alvarez,
  K.~Arnold, P.~Ashton, J.~Austermann, {\em et~al.}, ``The simons observatory:
  science goals and forecasts,'' {\em Journal of Cosmology and Astroparticle
  Physics}, vol.~2019, pp.~056--056, feb 2019.

\bibitem{cmbs4}
K.~Abazajian, G.~E. Addison, P.~Adshead, Z.~Ahmed, D.~Akerib, A.~Ali, S.~W.
  Allen, D.~Alonso, M.~Alvarez, M.~A. Amin, {\em et~al.}, ``{CMB}-s4:
  Forecasting constraints on primordial gravitational waves,'' {\em The
  Astrophysical Journal}, vol.~926, p.~54, feb 2022.

\bibitem{lsst_whitepaper}
J.~A. Tyson, ``{Large Synoptic Survey Telescope: overview},'' in {\em Survey
  and Other Telescope Technologies and Discoveries} (J.~A. Tyson and S.~Wolff,
  eds.), vol.~4836, pp.~10 -- 20, International Society for Optics and
  Photonics, SPIE, 2002.

\bibitem{lsst_srd}
Z.~Ivezi\'{c} and the LSST Science~Collaboration, ``Large synoptic survey
  telescope science requirements document,'' 2011.

\bibitem{euclid_whitepaper}
R.~Laureijs, J.~Amiaux, S.~Arduini, J.~L. Auguères, J.~Brinchmann, R.~Cole,
  M.~Cropper, C.~Dabin, L.~Duvet, A.~Ealet, {\em et~al.}, ``Euclid definition
  study report,'' 2011.

\bibitem{roman_whitepaper}
R.~Akeson, L.~Armus, E.~Bachelet, V.~Bailey, L.~Bartusek, A.~Bellini,
  D.~Benford, D.~Bennett, A.~Bhattacharya, and R.~B. andothers, ``The wide
  field infrared survey telescope: 100 hubbles for the 2020s,'' 2019.

\bibitem{desc_srd}
{The LSST Dark Energy Science Collaboration}, R.~Mandelbaum, T.~Eifler,
  R.~Hložek, T.~Collett, E.~Gawiser, D.~Scolnic, D.~Alonso, H.~Awan,
  R.~Biswas, {\em et~al.}, ``The lsst dark energy science collaboration (desc)
  science requirements document,'' 2021.

\bibitem{lsst_mg}
A.~Fert{\'{e}}, D.~Kirk, A.~R. Liddle, and J.~Zuntz, ``Testing gravity on
  cosmological scales with cosmic shear, cosmic microwave background
  anisotropies, and redshift-space distortions,'' {\em Physical Review D},
  vol.~99, apr 2019.

\bibitem{euclid_forecast}
{Euclid Collaboration}, A.~{Blanchard}, S.~{Camera}, C.~{Carbone}, V.~F.
  {Cardone}, S.~{Casas}, S.~{Clesse}, S.~{Ili{\'c}}, M.~{Kilbinger},
  T.~{Kitching}, M.~{Kunz}, {\em et~al.}, ``{Euclid preparation. VII. Forecast
  validation for Euclid cosmological probes},'' {\em A\&A}, vol.~642, p.~A191,
  Oct. 2020.

\bibitem{roman_forecast}
T.~Eifler, H.~Miyatake, E.~Krause, C.~Heinrich, V.~Miranda, C.~Hirata, J.~Xu,
  S.~Hemmati, M.~Simet, P.~Capak, {\em et~al.}, ``{Cosmology with the Roman
  Space Telescope – multiprobe strategies},'' {\em Monthly Notices of the
  Royal Astronomical Society}, vol.~507, pp.~1746--1761, 07 2021.

\bibitem{desy3}
T.~Abbott, M.~Aguena, A.~Alarcon, S.~Allam, O.~Alves, A.~Amon,
  F.~Andrade-Oliveira, J.~Annis, S.~Avila, D.~Bacon, {\em et~al.}, ``Dark
  energy survey year 3 results: Cosmological constraints from galaxy clustering
  and weak lensing,'' {\em Physical Review D}, vol.~105, jan 2022.

\bibitem{hsc32pt}
S.~More, S.~Sugiyama, H.~Miyatake, M.~M. Rau, M.~Shirasaki, X.~Li, A.~J.
  Nishizawa, K.~Osato, T.~Zhang, M.~Takada, {\em et~al.}, ``Hyper suprime-cam
  year 3 results: Measurements of clustering of sdss-boss galaxies,
  galaxy-galaxy lensing and cosmic shear,'' 2023.

\bibitem{kids1000}
C.~Heymans, T.~Tröster, M.~Asgari, C.~Blake, H.~Hildebrandt, B.~Joachimi,
  K.~Kuijken, C.-A. Lin, A.~G. S{\'{a} }nchez, J.~L. van~den Busch, {\em
  et~al.}, ``{KiDS}-1000 cosmology: Multi-probe weak gravitational lensing and
  spectroscopic galaxy clustering constraints,'' {\em Astronomy {\&}
  Astrophysics}, vol.~646, p.~A140, feb 2021.

\bibitem{stage4}
A.~Albrecht, G.~Bernstein, R.~Cahn, W.~L. Freedman, J.~Hewitt, W.~Hu, J.~Huth,
  M.~Kamionkowski, E.~W. Kolb, L.~Knox, J.~C. Mather, S.~Staggs, and N.~B.
  Suntzeff, ``Report of the dark energy task force,'' 2006.

\bibitem{fnl_forecast}
T.~Giannantonio, C.~Porciani, J.~Carron, A.~Amara, and A.~Pillepich,
  ``Constraining primordial non-gaussianity with future galaxy surveys,'' {\em
  Monthly Notices of the Royal Astronomical Society}, vol.~422, pp.~2854--2877,
  apr 2012.

\bibitem{fnl_forecast_euclidska}
D.~Yamauchi, K.~Takahashi, and M.~Oguri, ``Constraining primordial
  non-gaussianity via a multitracer technique with surveys by euclid and the
  square kilometre array,'' {\em Phys. Rev. D}, vol.~90, p.~083520, Oct 2014.

\bibitem{spherex}
O.~Doré, M.~W. Werner, M.~L.~N. Ashby, L.~E. Bleem, J.~Bock, J.~Burt,
  P.~Capak, T.-C. Chang, J.~Chaves-Montero, C.~H. Chen, {\em et~al.}, ``Science
  impacts of the spherex all-sky optical to near-infrared spectral survey ii:
  Report of a community workshop on the scientific synergies between the
  spherex survey and other astronomy observatories,'' 2018.

\bibitem{fnl_shear}
S.~{Hilbert}, L.~{Marian}, R.~E. {Smith}, and V.~{Desjacques}, ``{Measuring
  primordial non-Gaussianity with weak lensing surveys},'' {\em \mnras},
  vol.~426, pp.~2870--2888, Nov. 2012.

\bibitem{fnl_ia}
F.~Schmidt, N.~E. Chisari, and C.~Dvorkin, ``Imprint of inflation on galaxy
  shape correlations,'' {\em Journal of Cosmology and Astroparticle Physics},
  vol.~2015, p.~032, oct 2015.

\bibitem{Gatti_2020}
M.~Gatti, C.~Chang, O.~Friedrich, B.~Jain, D.~Bacon, M.~Crocce, J.~DeRose,
  I.~Ferrero, P.~Fosalba, and E.~G. andothers, ``Dark energy survey year 3
  results: cosmology with moments of weak lensing mass maps {\textendash}
  validation on simulations,'' {\em Monthly Notices of the Royal Astronomical
  Society}, vol.~498, pp.~4060--4087, aug 2020.

\bibitem{Prat_2023}
J.~Prat, J.~Zuntz, C.~Chang, T.~Tröster, E.~Pedersen, C.~Garc{\'{\i}
  }a-Garc{\'{\i}}a, E.~Phillips-Longley, J.~Sanchez, D.~Alonso, X.~Fang, {\em
  et~al.}, ``The catalog-to-cosmology framework for weak lensing and galaxy
  clustering for {LSST},'' {\em The Open Journal of Astrophysics}, vol.~6, apr
  2023.

\bibitem{Tan_2023}
T.~Tan, D.~Zürcher, J.~Fluri, A.~Refregier, F.~Tarsitano, and T.~Kacprzak,
  ``Assessing theoretical uncertainties for cosmological constraints from weak
  lensing surveys,'' {\em Monthly Notices of the Royal Astronomical Society},
  vol.~522, pp.~3766--3783, apr 2023.

\bibitem{Chandra_2022}
D.~Chandra and S.~Pal, ``Investigating the constraints on primordial features
  with future cosmic microwave background and galaxy surveys,'' {\em Journal of
  Cosmology and Astroparticle Physics}, vol.~2022, p.~024, sep 2022.

\bibitem{Huang_2012}
Z.~Huang, L.~Verde, and F.~Vernizzi, ``Constraining inflation with future
  galaxy redshift surveys,'' {\em Journal of Cosmology and Astroparticle
  Physics}, vol.~2012, pp.~005--005, apr 2012.

\bibitem{desy3_sr}
C.~S{\'{a} }nchez, J.~Prat, G.~Zacharegkas, S.~Pandey, E.~Baxter, G.~Bernstein,
  J.~Blazek, R.~Cawthon, C.~Chang, E.~Krause, {\em et~al.}, ``Dark energy
  survey year 3 results: Exploiting small-scale information with lensing shear
  ratios,'' {\em Physical Review D}, vol.~105, apr 2022.

\bibitem{Lewis:2002ah}
A.~Lewis and S.~Bridle, ``{Cosmological parameters from CMB and other data: A
  Monte Carlo approach},'' {\em \prd}, vol.~66, p.~103511, 2002.

\bibitem{Lewis:1999bs}
A.~Lewis, A.~Challinor, and A.~Lasenby, ``{Efficient computation of CMB
  anisotropies in closed FRW models},'' {\em \apj}, vol.~538, pp.~473--476,
  2000.

\bibitem{Howlett:2012mh}
C.~Howlett, A.~Lewis, A.~Hall, and A.~Challinor, ``{CMB power spectrum
  parameter degeneracies in the era of precision cosmology},'' {\em \jcap},
  vol.~1204, p.~027, 2012.

\bibitem{camb_notes}
A.~Lewis, ``{CAMB Notes}.'' \url{https://cosmologist.info/notes/CAMB.pdf}.

\bibitem{Mead_2020}
A.~J. Mead, S.~Brieden, T.~Tröster, and C.~Heymans, ``Hmcode-2020: improved
  modelling of non-linear cosmological power spectra with baryonic feedback,''
  {\em Monthly Notices of the Royal Astronomical Society}, vol.~502,
  pp.~1401--1422, jan 2021.

\bibitem{Bridle_2007}
S.~Bridle and L.~King, ``Dark energy constraints from cosmic shear power
  spectra: impact of intrinsic alignments on photometric redshift
  requirements,'' {\em New Journal of Physics}, vol.~9, pp.~444--444, dec 2007.

\bibitem{desy1_methods}
E.~Krause, T.~F. Eifler, J.~Zuntz, O.~Friedrich, M.~A. Troxel, S.~Dodelson,
  J.~Blazek, L.~F. Secco, N.~MacCrann, E.~Baxter, {\em et~al.}, ``Dark energy
  survey year 1 results: Multi-probe methodology and simulated likelihood
  analyses,'' 2017.

\bibitem{Zuntz_2015}
J.~Zuntz, M.~Paterno, E.~Jennings, D.~Rudd, A.~Manzotti, S.~Dodelson,
  S.~Bridle, S.~Sehrish, and J.~Kowalkowski, ``{CosmoSIS}: Modular cosmological
  parameter estimation,'' {\em Astronomy and Computing}, vol.~12, pp.~45--59,
  sep 2015.

\bibitem{lange2023nautilus}
J.~U. Lange, ``Nautilus: boosting bayesian importance nested sampling with deep
  learning,'' 2023.

\bibitem{mgs_bao}
A.~J. Ross, L.~Samushia, C.~Howlett, W.~J. Percival, A.~Burden, and M.~Manera,
  ``The clustering of the {SDSS} {DR}7 main galaxy sample {\textendash} i. a
  4~per cent distance measure at z~=~0.15,'' {\em Monthly Notices of the Royal
  Astronomical Society}, vol.~449, pp.~835--847, mar 2015.

\bibitem{boss_dr12}
S.~Alam, M.~Ata, S.~Bailey, F.~Beutler, D.~Bizyaev, J.~A. Blazek, A.~S. Bolton,
  J.~R. Brownstein, A.~Burden, C.-H. Chuang, {\em et~al.}, ``The clustering of
  galaxies in the completed {SDSS}-{III} baryon oscillation spectroscopic
  survey: cosmological analysis of the {DR}12 galaxy sample,'' {\em Monthly
  Notices of the Royal Astronomical Society}, vol.~470, pp.~2617--2652, mar
  2017.

\bibitem{sdss}
S.~Alam, M.~Aubert, S.~Avila, C.~Balland, J.~E. Bautista, M.~A. Bershady,
  D.~Bizyaev, M.~R. Blanton, A.~S. Bolton, J.~Bovy, {\em et~al.}, ``Completed
  {SDSS}-{IV} extended baryon oscillation spectroscopic survey: Cosmological
  implications from two decades of spectroscopic surveys at the apache point
  observatory,'' {\em Physical Review D}, vol.~103, apr 2021.

\bibitem{eboss_lrg1}
J.~E. {Bautista}, R.~{Paviot}, M.~{Vargas Maga{\~n}a}, S.~{de la Torre},
  S.~{Fromenteau}, H.~{Gil-Mar{\'\i}n}, A.~J. {Ross}, E.~{Burtin}, K.~S.
  {Dawson}, J.~{Hou}, {\em et~al.}, ``{The completed SDSS-IV extended Baryon
  Oscillation Spectroscopic Survey: measurement of the BAO and growth rate of
  structure of the luminous red galaxy sample from the anisotropic correlation
  function between redshifts 0.6 and 1},'' {\em \mnras}, vol.~500,
  pp.~736--762, Jan. 2021.

\bibitem{eboss_lrg2}
H.~Gil-Marín, J.~E. Bautista, R.~Paviot, M.~Vargas-Magaña, S.~de la Torre,
  S.~Fromenteau, S.~Alam, S.~Ávila, E.~Burtin, C.-H. Chuang, {\em et~al.},
  ``{The Completed SDSS-IV extended Baryon Oscillation Spectroscopic Survey:
  measurement of the BAO and growth rate of structure of the luminous red
  galaxy sample from the anisotropic power spectrum between redshifts 0.6 and
  1.0},'' {\em Monthly Notices of the Royal Astronomical Society}, vol.~498,
  pp.~2492--2531, 08 2020.

\bibitem{elg}
A.~de~Mattia, V.~Ruhlmann-Kleider, A.~Raichoor, A.~J. Ross, A.~Tamone, C.~Zhao,
  S.~Alam, S.~Avila, E.~Burtin, J.~Bautista, {\em et~al.}, ``The completed
  {SDSS}-{IV} extended baryon oscillation spectroscopic survey: measurement of
  the {BAO} and growth rate of structure of the emission line galaxy sample
  from the anisotropic power spectrum between redshift 0.6 and 1.1,'' {\em
  Monthly Notices of the Royal Astronomical Society}, dec 2020.

\bibitem{qso1}
R.~Neveux, E.~Burtin, A.~de Mattia, A.~Smith, A.~J. Ross, J.~Hou, J.~Bautista,
  J.~Brinkmann, C.-H. Chuang, K.~S. Dawson, {\em et~al.}, ``{The completed
  SDSS-IV extended Baryon Oscillation Spectroscopic Survey: BAO and RSD
  measurements from the anisotropic power spectrum of the quasar sample between
  redshift 0.8 and 2.2},'' {\em Monthly Notices of the Royal Astronomical
  Society}, vol.~499, pp.~210--229, 09 2020.

\bibitem{qso2}
J.~Hou, A.~G. S{\'{a} }nchez, A.~J. Ross, A.~Smith, R.~Neveux, J.~Bautista,
  E.~Burtin, C.~Zhao, R.~Scoccimarro, K.~S. Dawson, {\em et~al.}, ``The
  completed {SDSS}-{IV} extended baryon oscillation spectroscopic survey: {BAO}
  and {RSD} measurements from anisotropic clustering analysis of the quasar
  sample in configuration space between redshift 0.8 and 2.2,'' {\em Monthly
  Notices of the Royal Astronomical Society}, vol.~500, pp.~1201--1221, oct
  2020.

\bibitem{lyalpha}
H.~{du Mas des Bourboux}, J.~{Rich}, A.~{Font-Ribera}, V.~{de Sainte Agathe},
  J.~{Farr}, T.~{Etourneau}, J.-M. {Le Goff}, A.~{Cuceu}, C.~{Balland}, J.~E.
  {Bautista}, {\em et~al.}, ``{The Completed SDSS-IV Extended Baryon
  Oscillation Spectroscopic Survey: Baryon Acoustic Oscillations with
  Ly{\ensuremath{\alpha}} Forests},'' {\em \apj}, vol.~901, p.~153, Oct. 2020.

\bibitem{getdist}
A.~Lewis, ``{GetDist: a Python package for analysing Monte Carlo samples},''
  2019.

\bibitem{story_measurement_2013}
K.~T. Story, C.~L. Reichardt, Z.~Hou, R.~Keisler, K.~A. Aird, B.~A. Benson,
  L.~E. Bleem, J.~E. Carlstrom, C.~L. Chang, H.-M. Cho, {\em et~al.}, ``A
  {Measurement} of the {Cosmic} {Microwave} {Background} {Damping} {Tail} from
  the 2500-square-degree {SPT}-{SZ} survey,'' {\em The Astrophysical Journal},
  vol.~779, p.~86, Dec. 2013.
\newblock arXiv:1210.7231 [astro-ph].

\bibitem{Hinshaw_2013}
G.~Hinshaw, D.~Larson, E.~Komatsu, D.~N. Spergel, C.~L. Bennett, J.~Dunkley,
  M.~R. Nolta, M.~Halpern, R.~S. Hill, N.~Odegard, {\em et~al.}, ``Nine-{Year}
  {Wilkinson} {Microwave} {Anisotropy} {Probe} ({WMAP}) {Observations}:
  {Cosmological} {Parameter} {Results},'' {\em The Astrophysical Journal
  Supplement Series}, vol.~208, p.~19, Sept. 2013.

\bibitem{r2model1}
A.~A. {Starobinskij}, ``{Spectrum of adiabatic perturbations in the universe
  when there are singularities in the inflationary potential.},'' {\em Soviet
  Journal of Experimental and Theoretical Physics Letters}, vol.~55,
  pp.~489--494, May 1992.

\bibitem{r2model2}
A.~A. {Starobinsky}, ``{A new type of isotropic cosmological models without
  singularity},'' {\em Physics Letters B}, vol.~91, pp.~99--102, Mar. 1980.

\bibitem{r2model3}
V.~F. {Mukhanov} and G.~V. {Chibisov}, ``{Quantum fluctuations and a
  nonsingular universe},'' {\em Soviet Journal of Experimental and Theoretical
  Physics Letters}, vol.~33, p.~532, May 1981.

\end{thebibliography}

\end{document}